\documentclass[twocolumn,trackchanges]{aastex631}

\usepackage{booktabs}
\usepackage{multirow}
\DeclareUnicodeCharacter{2212}{-}

\begin{document}

\title{Exploring Unusual High-frequency Eclipses in MSP J1908$+$2105}

\author[0009-0002-3211-4865]{Ankita Ghosh}
\affiliation{National Centre For Radio Astrophysics, Tata Institute of Fundamental Research, Pune 411007, India}

\author[0000-0002-6287-6900]{Bhaswati Bhattacharyya}
\affiliation{National Centre For Radio Astrophysics, Tata Institute of Fundamental Research, Pune 411007, India}

\author[0000-0002-3764-9204]{Sangita Kumari}
\affiliation{National Centre For Radio Astrophysics, Tata Institute of Fundamental Research, Pune 411007, India}

\author[0000-0002-7122-4963]{Simon Johnston}
\affiliation{CSIRO Astronomy and Space Science, P.O. Box 76, Epping, NSW 1710, Australia}

\author[0000-0003-2122-4540]{Patrick Weltevrede}
\affiliation{Jodrell Bank Centre for Astrophysics, Department of Physics and Astronomy, The University of Manchester, UK}

\author[0000-0002-2892-8025]{Jayanta Roy}
\affiliation{National Centre For Radio Astrophysics, Tata Institute of Fundamental Research, Pune 411007, India}

\begin{abstract}
This paper presents a comprehensive study of the eclipse properties of the spider millisecond pulsar (MSP) J1908$+$2105, using wide-band observations from the uGMRT and Parkes UWL. For the first time, we observed that this pulsar exhibits extended eclipses up to 4 GHz, the highest frequency band of the UWL, making it one of only three MSPs known to have such high-frequency eclipses. This study reveals synchrotron absorption as the primary eclipse mechanism for J1908$+$2105. We present modeling of synchrotron optical depth with various possible combinations of the parameters to explain the observed eclipsing in this as well as other spider MSPs. Observed eclipses at unusually high frequencies for J1908$+$2105 significantly aided in constraining the magnetic field and electron column density in the eclipse medium while modeling the synchrotron optical depth. Combining our findings with data from other MSPs in the literature, for the first time we note that a higher cutoff frequency of eclipsing, particularly above 1 GHz, is consistently associated with a higher electron column density ($>$ 10$^{17}$ cm$^{-2}$) in the eclipse medium. Additionally, we present the first evidence of lensing effects near eclipse boundaries in this MSP, leading to significant magnification of radio emissions. The orbital phase resolved polarization analysis presented in this paper further indicates variation in rotation measure and consequently stronger magnetic fields in the eclipse region.

\end{abstract}


\section{Introduction} \label{sec:intro}

Millisecond pulsars (MSPs) in compact binary orbits having a low-mass non-degenerate or partially non-degenerate companion with an orbital period of less than a day are known as ``spider MSPs'' \citep{2013IAUS..291..127R}. Depending on the mass of the companion ($M_{c}$), these spider MSPs are further divided into two categories; ``black widow (BW)'' ($M_{c}$ $<$ 0.05 $M_{\odot}$) and ``redback (RB)''(0.1 $M_{\odot} < M_{c} <$ 0.9 $M_{\odot}$) spiders. As the pulsar and the companion are in very close proximity, in these systems, the highly energetic wind from the pulsar ablates the companion. Material blown from the companion causes an eclipse by obscuring the pulsar's radio emission. Frequency-dependent eclipsing is reported for a majority of known spider MSPs. Such eclipsing MSP systems can aid in the understanding of properties of the low-mass companions in tight binary orbits, plasma properties of eclipse material, mass flow from the companion driven by relativistic pulsar wind, orbital properties in strong gravitational potential, etc. \cite{1994ApJ...422..304T} reviewed a set of plausible eclipse mechanisms suggested by several authors. The majority of these works proposed cyclo-synchrotron absorption as a likely eclipse mechanism for low-frequency eclipses and scattering/stimulated Raman scattering as the cause of high-frequency eclipses.  Understanding the eclipse mechanism allows us to investigate the subsequent properties of eclipsing materials as well as the interpretation of the intrabinary shock that happens in these systems as a result of stellar wind and relativistic pulsar wind interaction. 

Polarization properties are closely tied to the local environment of a pulsar, making polarization studies of the eclipse medium in spider MSPs crucial for uncovering the mysteries of the eclipse mechanism \citep{2019MNRAS.490..889P}. Spider MSPs are particularly intriguing because tracking the evolution of rotational measure (RM) and polarization position angle (PPA) swings throughout the orbit can provide insights into the dynamic state of the ablated material in the eclipse medium. Analyzing the polarization of these pulsars reveals a substantial magnetic field within the eclipse medium, suggesting possible cyclo-synchrotron absorption. Recent polarization studies and plasma lensing predictions at eclipse ingress and egress \citep{https://doi.org/10.1093/mnras/stz374, 2019MNRAS.490..889P, https://doi.org/10.1093/mnras/staa933} support this, with \citet{https://doi.org/10.1093/mnras/stz374} showing how magnetic fields cause different magnifications in time and frequency for the two circular polarizations.

Motivated by this, we present the eclipse and polarization properties of spider pulsar PSR J1908$+$2105 using the observations from the Giant Meter Radio Telescope (GMRT) and Parkes radio telescope.
PSR J1908$+$2105 was discovered by searching for radio pulsations targetted towards the unidentified Fermi-Large Area Telescope (LAT) source with the 327 MHz receiver of the Arecibo \citep{2016ApJ...819...34C}. It has a spin period of 2.56 ms and a dispersion measure (DM) of 61.9067 pc/cm$^{-3}$, with a flux density of 0.6 mJy at 327 MHz. \cite{2021ApJ...909....6D} performed a timing analysis of this spider pulsar, revealing that it has an orbital period of 3.51 hours and a minimum companion mass of 0.055 $M_{\odot}$, placing it between black widows and redbacks in the orbital period versus companion mass space. PSR J1908$+$2105 also exhibits unique characteristics sharing properties from both black widows and redback spiders similar to another such system J1242$-$4712 \citep{http://dx.doi.org/10.3847/1538-4357/ad31ab} suggesting a category of objects that share properties bridging these two subclasses of spider binary systems. Given the rarity of such systems, a detailed study is essential to explore the potential evolutionary transition from redbacks to black widows. For J1908$+$2105, the system's short orbital period and relatively small companion mass align with characteristics typical of black widow pulsars, whereas the observed extended eclipses are typical characteristics of redback pulsars. \cite{2021ApJ...909....6D} suggested the presence of an unusual companion size or a dense plasma distribution surrounding the companion, as the possible reason for the extended eclipse in PSR J1908$+$2105. 
The eclipse mechanism and polarization properties of this system have not been explored. In this paper, we aim to constrain the magnetic field in the eclipse medium by thoroughly investigating the eclipse properties using wide-bandwidth polar observations. 

Our observations and analysis methods for this paper are described in
Section \ref{sec:observation}. In Section \ref{sec:Results} we investigate how the pulsar’s eclipse and polarization properties vary as functions of orbital phase. We discuss the possible eclipse eclipse mechanism in Section \ref{sec:eclipse mec} and present the conclusions in Section \ref{sec:summary}.

\section{Observations and data analysis} \label{sec:observation}

\renewcommand{\tabcolsep}{2.5pt}
\begin{table}[!htb]
\begin{center}
\caption{Summary of observations}
\footnotesize{
\label{tab:1}
\begin{tabular}{ccccccc}
%\hline
\toprule
Backend & $T^{a}_{res}$  &  $F^{b}_{res}$  & Bandwidth  &
  Centre frequency & No. of \\
  & ($\mu s $) & (MHz) & (MHz) & (MHz) & epochs \\ \hline
Parkes UWL & 5.189 &  1.0 & 3328 &  2368 & 2 \\ \hline
uGMRT & 10.24 & 0.39 & 200 & 400 & 3\\ \hline
uGMRT & 10.24 & 0.39 & 200& 650 & 2 \\ \bottomrule

\hline
\end{tabular}
}
\end{center}
{\footnotesize {\bf{Notes.}}\\
$^{a}$: Time resolution\\
$^{b}$: Frequency resolution\\}
\end{table} 

We observed PSR J1908$+$2105 with the uGMRT  \citep{1991ASPC...19..376S, 2017CSci..113..707G, http://dx.doi.org/10.1142/S2251171716410117} in band$-$3 (300$−$500 MHz) and band$-$4 (550$−$750 MHz) in several epochs as listed in Table \ref{tab:2}. The data were coherently dedispersed across each frequency sub-band of 390 kHz using  DM of 61.9067 pc/cm$^{-3}$. We recorded data at a rate of 48 MB/s, employing 8-bit sampling, and used 512 channels with a sampling time of 10.24 $\mu$s.  To effectively mitigate the impact of narrowband and short-duration broadband radio frequency interference (RFI), we used real-time RFI filter \citep{2019JAI.....840006B,2022JAI....1150008B} as well as RFI mitigation software in conjunction with the GMRT pulsar tool (gptool\footnote{\url{https://github.com/chowdhuryaditya/gptool}}).
To eliminate inter-channel smearing, we performed incoherent dedispersion to the obtained filterbanks.
Then to obtain the folded profile from each epoch's observations, we used the {\sc{PRESTO}} task ``{\sc{prepfold}}'', to fold each data set using the parameter file of the pulsar.\par

We observed PSR J1908$+$2105 using the Parkes Ultra-Wideband (UWL) receiver \citep{2020PASA...37...12H} in two epochs 2023 October 21 and 31. The observational frequency range spans from 704 to 4032 MHz, divided into 3328 channels with a frequency resolution of 1 MHz. The data were then coherently de-dispersed and folded using the topocentric periodicity of the pulsar. The folded data were organized into subintegrations of duration of 30 seconds with 1024 phase bins per pulsar period. Observations have a duration of approx 2.5 hours covering nearly one full orbit of the pulsar. Table \ref{tab:1} provides a summary of these observations.

Before observing the target source, we briefly monitored a pulsed calibration signal injected into the low-noise amplifiers. We used the pipeline {\sc PSRPYPE}\footnote {\url{https://github.com/vivekvenkris/psrpype/}} for RFI mitigation. The remaining RFI was manually mitigated in the time and frequency domain. {\sc PSRCHIVE} \citep{2004PASA...21..302H} was used for the data processing. The routine “{\sc pac}” corrects for instrumental gain and phase differences using the pulsed calibrator, carries out flux calibration obtained from observations of PKS B1934$–$68 and corrects for instrumental leakage terms using the PCM \citep{2004ApJS..152..129V} method. To improve the signal-to-noise ratio (SNR), we averaged across frequency and phase bins by factors of 8. This end result is a calibrated folded profile comprising 4 Stokes parameters, 416 frequency channels, and 128 phase bins for each epoch.
Consecutive observations excluding the eclipse region were combined using the task “{\sc psradd}”, generating a single output file. A time drift observed in the combined file was rectified by updating the pulsar ephemeris. This involved fitting the time of arrivals (TOAs) calculated using ``{\sc pat}”, solely for spin frequency (F0), the epoch of ascending node (T0), and projected semi-major axis (A1) using {\sc TEMPO2} \citep{2006MNRAS.369..655H}. We used this combined average profile from the non-eclipse phase to obtain the average polarization fraction and rotational measure of the pulsar.

\section{Results and Interpretations} \label{sec:Results}

\subsection{Profile evolution} \label{Profile evolution}

\begin{figure}[ht!]
\begin{center}
\includegraphics[width=0.4\textwidth,angle=0]{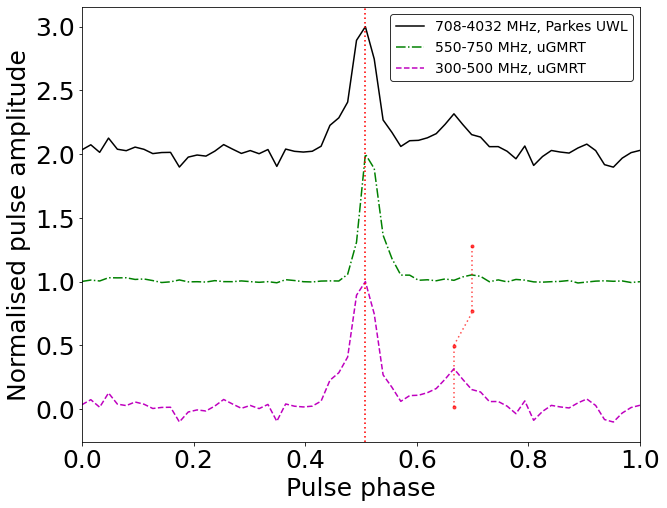}
\caption{Normalized averaged pulse profiles in the non-eclipse region from coherently de-dispersed GMRT observation at 400 MHz (dashed-magenta line) and 650 MHz (dash-dotted green line) and Parkes UWL observations at 2368 MHz (solid black line). The vertical dashed line is a visual guide to show how the main peaks at various frequencies are intrinsically aligned. The inflections in the red dotted line indicate shifts in the relative position of the components at higher frequencies.}
\label{fig:prof-evol}
\end{center}
\end{figure}

PSR J1908$+$2105 shows significant profile evolution as we move towards the higher frequencies. A second component is observed to be present at 400 MHz which is absent at 2368 MHz (Figure \ref{fig:prof-evol}). The main component has a width at 50\% of the peak intensity ($W_{50}$) of approximately 0.10 $\pm$ 0.02 ms at a frequency of 2368 MHz and 0.17$\pm$0.04 ms at 400 MHz, and the $W_{50}$ of the second component is 0.19 $\pm$ 0.04 ms at 400 MHz. In the non-eclipse phase, the scattering time scale is calculated to be $\tau$ = 0.08 $\pm$ 0.04 ms and 0.05 $\pm$ 0.02 ms at 400 MHz and 1200 MHz respectively by fitting the pulse profile with a convolution of Gaussian and exponential functions assuming the thin screen model. Whereas, the calculated scattering time scale is found to be 0.001 ms at 1200 MHz and 0.06 ms at 400 MHz using the DM model given by \cite{2004ApJ...605..759B}.

\subsection{Frequency dependent eclipse} \label{frequency dependence}
In most spider systems, the duration of the eclipse decreases with increasing frequency, and eclipses often disappear entirely above 1.4 GHz. PSR J1908$+$2105 eclipses for $\sim$ 40\%  of the orbit at 327 MHz \citep{2021ApJ...909....6D}. We found that the eclipse duration is approximately 38\% at 400 MHz, reducing to 31\% at 650 MHz. At 2368 MHz, we couldn't precisely probe the egress boundary where the pulse reappears after the eclipse. Assuming a symmetric eclipse and that the eclipse center does not change from our observations on 16 Dec 2023, we estimate the duration to be around 30\% at this frequency. However, as shown in Table \ref{tab:2} and Figure \ref{fig:egress}, the center of the eclipse at 650 MHz is shifted to the left compared to 400 MHz. This shift may be a temporal effect due to the turbulent nature of the eclipse medium, given the considerable time gap between the observations.

\subsection{Electron density distribution near eclipse} \label{eclipse obs}

\begin{figure*}[ht!]
\begin{center}
\includegraphics[width=0.8\textwidth,angle=0]{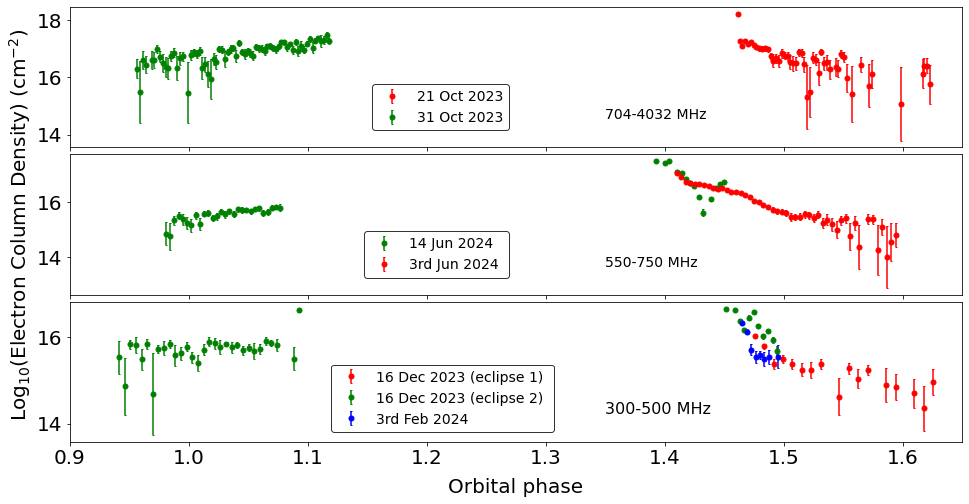}
\caption{Electron density variation near eclipse boundary in different epochs observed with uGMRT at 400 MHz; 650 MHz and with Parkes UWL at 2368 MHz.  A sharp dip around orbital phase \( \sim 1.42 \) on 14 June 2024 suggests a localized gap in the ionized gas, likely caused by turbulence in the plume. The absence of this dip in data from 11 days earlier indicates variability in timescales of days to weeks.}
\label{fig:egress}
\end{center}
\end{figure*}

PSR J1908$+$2105 shows an eclipse for approximately 38\% of its orbit at 400 MHz. 
We investigated the orbital phase-dependent DM variation for PSR J1908$+$2105 and observed an increase in the DM value near the eclipse boundaries. The excess DM, and the corresponding electron column density (N$_{e}$) as the function of the orbital phase are (illustrated in Figure \ref{fig:egress}) determined from the delay in the time of arrival of pulses (obtained using {\sc TEMPO2}) using the relation,
\begin{equation}
    \label{eq:dm_ex}
    \text{DM}_{ex} \;(pc\; cm^{-3})=2.4 \times 10^{-10}\times \text{TOA delay}(\mu s)\times f^{2} (MHz)
\end{equation}
where TOA delay is the time delay in $\mu$s and f is the observing frequency in MHz. From DM$_{ex}$ the excess N$_{e}$ is computed using,
\begin{equation}
    \label{eq:ne_ex}
   \text{Excess}\;\; N_{e}\;(cm^{-2})=3 \times 10^{18}\times \text{DM}_{ex} (pc \; cm^{-3})
\end{equation}
We were able to observe the ingress (when the pulsar is entering the eclipse) of the eclipse only during one epoch, while the egress (when the pulsar is coming out of the eclipse) was observed in three epochs using uGMRT band$-$3 observations. During the ingress of the MSP, the loss of pulsation is very abrupt, whereas, during eclipse egress, it gradually decreases, leaving a distinct trail of materials (Figure \ref{fig:egress}). Table \ref{tab:2} details temporal shifts of eclipse boundaries corresponding to different epochs. For example, The orbital phase at which the pulse reappears at 400 MHz was observed to vary between 0.447 and 0.485, translating to an approximately 8-minute difference. Such variability suggests that the boundary of ablated material around the companion is not sharply defined causing an unstable and turbulent egress. The eclipse boundaries are determined by observing the signal-to-noise ratio of sub-integrations near the eclipse region, specifically where the S/N falls below 4.
\renewcommand{\tabcolsep}{1.5pt}
\begin{table}[!htb]
\begin{center}
\caption{Temporal change of eclipse boundaries}
\footnotesize{
\label{tab:2}
\begin{tabular}{ccccccc}
%\hline
\toprule
Backend & Frequency  &  Date of  & Eclipse  & Orbital \\
 
  & (MHz) & observations & boundary  & phase ($\phi_{p}$) \\ \hline
\multirow{2}{*}{Parkes UWL}  & \multirow{2}{*}{2368} & 21 Oct 2023 & egress & $0.34<\phi_{p}<0.45$ \\
 &    & 31 oct 2023 & ingress &  0.125 $\pm$ 0.002  \\ \hline

 \multirow{2}{*}{uGMRT}  & \multirow{2}{*}{650} &  03 Jun 2024  & ingress & $0.012<\phi_{p}<0.085$ \\ 
 &    & 03 Jun 2024  & egress &  0.390 $\pm$ 0.002 \\
 &    & 14 Jun 2024 & ingress & 0.076 $\pm$ 0.002 \\
 &    & 14 Jun 2024  & egress &  0.387 $\pm$ 0.002  \\  \hline

\multirow{2}{*}{uGMRT}  & \multirow{2}{*}{400} & 16 Dec 2023 & egress & 0.475 $\pm$ 0.002 \\
 &    & 16 Dec 2023  & ingress &  0.098 $\pm$ 0.005  \\ 
 &    & 16 Dec 2023  & egress &  0.447 $\pm$ 0.004 \\ 
 &    & 04 Feb 2024  & egress &  0.465 $\pm$ 0.002 \\  \hline

\end{tabular}
}
\end{center}
{\footnotesize {\bf{Note.}}\\
The eclipse boundary was determined where the signal-to-noise ratio drops below 4$\sigma$.}
\end{table} 

 The e-folding time, as defined in \cite{1991ApJ...380..557R} for PSR B1957$+$20, is the time over which the local column density decreases by a factor of \( e \sim 2.718 \). At eclipse ingress for J1908$+$2105, we determined the change in orbital phase corresponding to an e-folding of the local column density to be, $\Delta\phi \sim$ 0.0047. During this time the companion star advances in its orbit by about 25,000 km. At eclipse egress, the e-folding length scale is about 3 times larger, corresponding to $\Delta\phi\sim$ 0.0157 (Figure \ref{fig:egress}).
During this time the companion star advances in its orbit by about 86,000 km. This explains the long trail observed near egress caused by the sweeping of the stellar material due to the companion's orbital motion as found by \cite{https://doi.org/10.1038/333237a0} for PSR B1957$+$20. The asymmetric distribution of material at either side of the eclipse is also seen for black widow PSR B1957$+$20 \citep{1991ApJ...381L..21T, 1991ApJ...380..557R}. As noted in \cite{2018MNRAS.476.1968P}, magnetic reconnection between the pulsar wind's magnetic field and the companion's magnetosphere can result in eclipse material leaking into the pulsar wind. This process may explain the excess material observed in the eclipse tail as the companion progresses in its orbit. However, whether the tail of excess material appears at egress or ingress depends on the combined influence of gravitational pressure, Coriolis forces, and the relative velocity of the pulsar wind compared to the companion's orbital velocity \citep{1991ApJ...381L..21T}. For instance, in the case of PSR J1744$−$24A, these effects push the eclipse material toward ingress, forming a tail near the ingress boundary, whereas for PSR B1957$+$20, the material is pushed toward the egress \citep{1991ApJ...381L..21T}. This aligns with the two-dimensional SPH (Smoothed-Particle Hydrodynamics) calculations of eclipse outflow by \cite{1991ApJ...381L..21T}. At 400 MHz, GMRT observations on 16 Dec 2023 show $\Delta$DM $\sim$ 0.02 pc cm$^{-3}$ near orbital phase $\phi = 0.48$ during egress. However, moving closer to the companion, at 2368 MHz, Parkes observations on 21 Oct 2023 show $\Delta$DM $\sim$ 0.6 pc cm$^{-3}$ near $\phi = 0.455$. Therefore, assuming, that the eclipse center did not change between the observations, we can calculate the slope of increase in DM with orbital phase ($\frac{\delta(\Delta \text{DM})}{\Delta \phi})\approx 23$ pc cm$^{-3}$. Similarly, during ingress, uGMRT observations show $\Delta$DM $\sim$ 0.013 pc cm$^{-3}$ near $\phi = 0.09$, increasing to 0.18 pc cm$^{-3}$ at 2368 MHz near $\phi = 0.125$ and therefore, the slope of increase in DM with orbital phase ($\frac{\delta(\Delta \text{DM})}{\Delta \phi})\approx 5$ pc cm$^{-3}$, indicating a sharp rise in DM at egress compared to ingress when we move closer towards the companion. Thus, we could predict the possible $\Delta$ DM $\sim$ 4 pc cm$^{-3}$ at the eclipse center ($\phi$ $\sim$ 0.28) and corresponding electron column density $(N_{e})$ to be $\approx 10^{19}$ cm$^{-2}$.
Despite the typical trend of eclipse duration decreasing with frequency, this MSP exhibits an unusually prolonged eclipse, covering over 30\% of the orbit even at frequencies up to 4 GHz, and likely extending well beyond. Such eclipses at high frequencies are exceedingly rare and, to our knowledge, have been reported for only two other spider MSPs (e.g. PSR J1723$-$2837 and J1731$-$1847). The eclipse mechanisms at such high frequencies have not been thoroughly investigated for either of these two MSPs.
This necessitates constraining the eclipse mechanism that can account for eclipses occurring even at such high frequencies.

\subsection{Orbital Phase resolved DM and polarization properties} \label{Polarization prop}

\begin{figure}[ht!]
\begin{center}
\includegraphics[width=0.4\textwidth,angle=0]{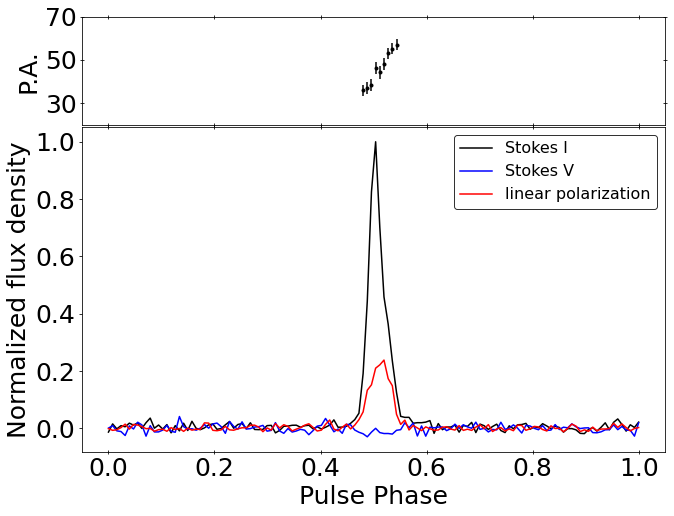}
\caption{Polarization profile at 2368 MHz from Parkes UWL observations in the non-eclipse region. The upper panel shows the PPA variations (in black dots) and RVM fit (in red line) to it.}
\label{fig:pol}
\end{center}
\end{figure}

\begin{figure*}[ht!]
\begin{center}
\includegraphics[width=0.8\textwidth,angle=0]{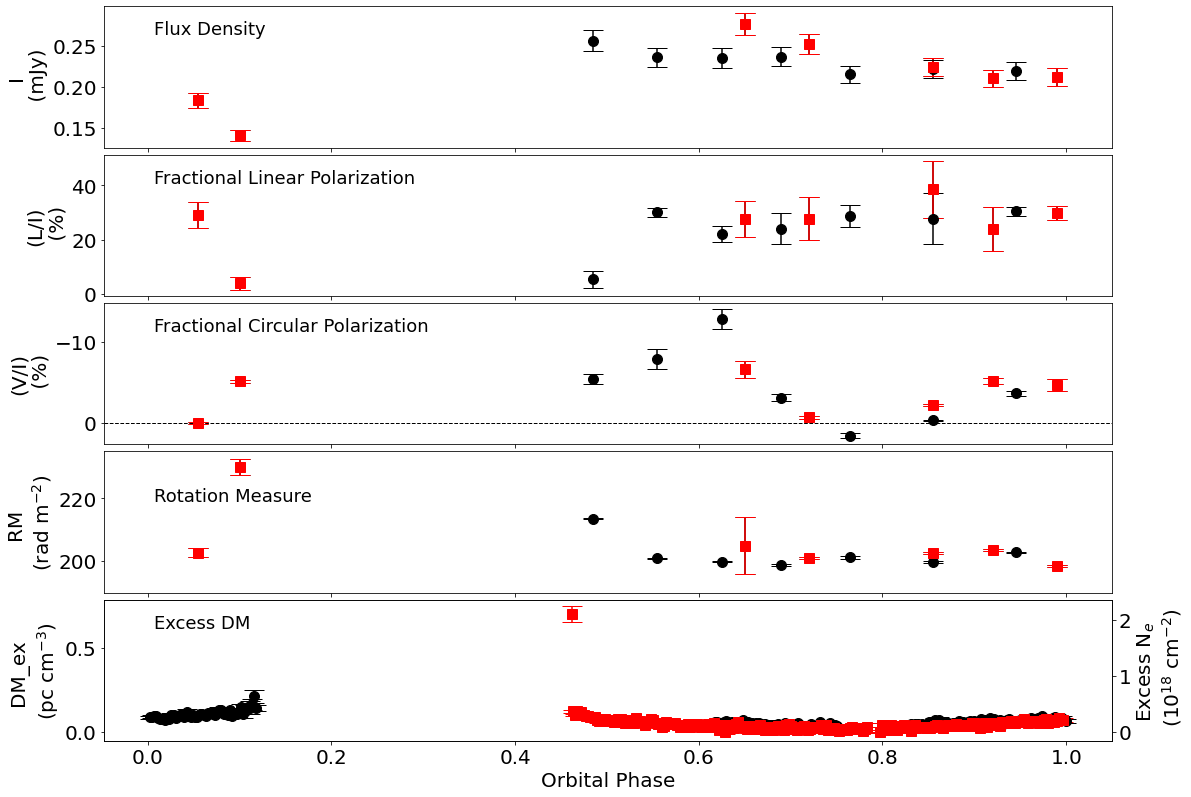}
\caption{Variation of total intensity, polarization fractions along with RM and DM as a function of orbital phase for PSR J1908$+$2105 at 2368 MHz from the Parkes UWL observations. Here, the different symbol in a single panel shows observations from different epochs.}
\label{fig:flux-pol}
\end{center}
\end{figure*}

 We have performed a polarization study on PSR J1908$+$2105 using our observations from Parkes UWL (708$-$4032 MHz) only. 
 We have obtained the integrated polarization profile (Figure \ref{fig:pol}) for J1908$+$2105 by adding the observations of the non-eclipse phase, which are calibrated and RM corrected. We report the best fit RM value is $201$ $\pm$ 1 rad/m$^{2}$ from the integrated average profile of the pulsar. The average pulse profile shows linear polarisation of nearly 32\% in the non-eclipse phase. The circular polarization fraction is comparably low $\sim$ 2.8\%. We used the full bandwidth of 3328 MHz of the frequency band (708$-$4032 MHz) to calculate the total intensity and polarization properties.
We analyzed the variation of intensity and polarization fraction, as well as RM and DM as a function of the orbital phase. Flux density calculations were done following the method explained in \cite{https://doi.org/10.3847/1538-4357/ac19b9}.  We used {\sc TEMPO2} to find the DM variation with the orbital phase. We observed the effect of the eclipse through a decrease in total intensity and linear polarization fraction and an increase in DM and RM around superior conjunction (orbital phase, 0.12$-$0.40) (Figure \ref{fig:flux-pol}). Where the average RM is $\sim$ 201 rad/m$^{2}$ for the non-eclipse phase, near the eclipse medium RM increases to 230 rad/m$^{2}$. The increase in DM and RM shows the increased electron density near the eclipse medium. The orbital phase-dependent flux density is calculated considering the full bandwidth of Parkes UWL band, with a central frequency of 2368 MHz and a bandwidth of 3328 MHz.\\
Depolarization during the eclipse phase has been observed in spider MSPs; for example, PSR J2051$−$0827 by \cite{https://doi.org/10.3847/1538-4357/acea81}, PSR J2256$−$1024 by \cite{https://doi.org/10.1093/mnras/staa933}, and J1748$−$2446A by \cite{2018ApJ...867...22Y}. This depolarization can occur due to the increased RM near eclipse as well as multipath propagation of pulsed radiation through the circumstellar magnetized plasma leading to rapid time variations in the RM. Such rapid RM variations can result from fluctuations in either or both of the circumstellar components of DM and the parallel component of the magnetic field. DM variations are observed at both eclipse ingress and egress for eclipsing spider pulsars. However, these variations are typically very low, which will not significantly affect the RM \citep{2018ApJ...867...22Y}. We investigated the orbital phase-dependent DM variation for PSR J1908$+$2105 and observed an increase in the DM value during the eclipse phase (around $\sim$0.12$-$0.40), however not very significant. The excess DM and the corresponding electron column density (N$_{e}$) as the function of the orbital phase is illustrated in Figure \ref{fig:flux-pol}. These maximum DM variations are found to be of the order $\sim$ 0.07 pc cm$^{−3}$ and 0.6 pc cm$^{−3}$ near the ingress and egress boundary respectively at the observing frequency of Parkes. Since RM can fluctuate with changes in DM and magnetic field, the small increase in DM observed is unlikely to significantly impact the RM. Therefore, short-timescale variations in the line-of-sight magnetic field can cause depolarization through rapid fluctuations of RM in time, indicating that there might be a significant magnetic field in the eclipse medium, which can also be inferred from the increased RM near the eclipse medium for this MSP. Such fluctuation of the magnetic field is common in the turbulent stellar wind.

%calculate ionospheric RM 

\subsection{Magnetic field in the eclipse medium} \label{Magnetic Field}

During the ingress of the eclipse, the maximum RM is about 230 rad m$^{−2}$ with a $\Delta$RM$_{max}$ of 30 rad m$^{−2}$ compared to that of the non-eclipse phase. The $\Delta$DM at the orbital phase of $\Delta$RM$_{max}$ is 0.05 pc cm$^{−3}$. The line-of-sight magnetic field strength in the eclipse boundary can be estimated by measuring the changes in the Faraday rotation, $B_{||}$=1.23$\mu$G$\frac{\Delta \text{RM}}{\Delta \text{DM}}$=0.9 mG.\par
Similarly, during the egress of the eclipse, the maximum RM is about 213 rad m$^{−2}$ with a $\Delta$RM$_{max}$ of 13 rad m$^{−2}$ compared to that of the non-eclipse phase. The $\Delta$DM at the orbital phase of $\Delta$RM$_{max}$ is 0.017 pc cm$^{−3}$. The line-of-sight magnetic field strength in the eclipse egress is therefore estimated to be $\sim$ 1 mG.
Since PSR J1908$+$2105 shows a complete eclipse till 4 GHz, we were not able to obtain the magnetic field in the eclipse center through RM variation. Therefore, for the center of eclipse medium, we calculated the characteristic magnetic field (B$_{E}$), using the pressure balance between pulsar wind energy density (U$_{E}$ = $\frac{\dot{E}}{4\pi c a^{2}}$ ) and the stellar wind energy density of the companion ($\frac{B_{E}}{8\pi}$), where a is the distance between the pulsar and the companion (a $\sim$ 1.30 R$_{\odot}$) and c is the speed of light. This gives the characteristic magnetic field for PSR J1908$+$2105 to be $\sim$ 16 G in the center of the eclipse medium.
There have been efforts to constrain the magnetic fields of several other spider pulsars.
For instance, \cite{https://doi.org/10.3847/1538-4357/acea81} estimated the magnetic field for J2051$−$0827 to be 0.1 G at the eclipse boundary from the observed RM variation. \cite{2019MNRAS.490..889P} analyzed the polarization properties of PSR J2051$−$0827, providing tentative constraints on the line-of-sight magnetic field (20 ± 120 G) and the (near-) perpendicular field ($<$ 0.3 G) inside the eclipse medium. \cite{https://doi.org/10.1093/mnras/staa933} derived the magnetic field for PSR J2256$−$1024 from RM variations near the eclipse boundary, suggesting a line-of-sight magnetic field strength of approximately 1.11 mG at the eclipse boundary. \citet{https://doi.org/10.1093/mnras/stz374} constrained the line-of-sight magnetic field ($B_{||}$) and its spatial structure ($\sigma$B) using plasma lensing, finding the line-of-sight magnetic field strength near the eclipse boundary of PSR B1957$+$20 to be less than 0.02 G. Additionally, \citet{https://doi.org/10.1093/mnras/stac3456} detected evidence of Faraday conversion and attenuation in PSR B1744$−$24A, estimating the magnetic field in the eclipse medium to be approximately 100 G.

\subsection{Constraining the emission geometry}
The polarization position angle (PPA) in the non-eclipse medium showed a hint of swing over the pulse profile. We have explored the Rotating Vector Model (RVM, defined in Eq. \ref{rvm}, \citet{1969ApL.....3..225R}) to fit the swing and constrain the emission geometry.

\begin{equation}
\label{rvm}
\tan(\psi - \psi_0) = \frac{\sin\alpha \sin(\phi - \phi_0)}{\sin(\alpha + \beta)\cos\alpha - \cos(\alpha + \beta)\sin\alpha \cos(\phi - \phi_0)}
\end{equation}

\begin{figure}[ht!]
\begin{center}
\includegraphics[width=0.4\textwidth,angle=0]{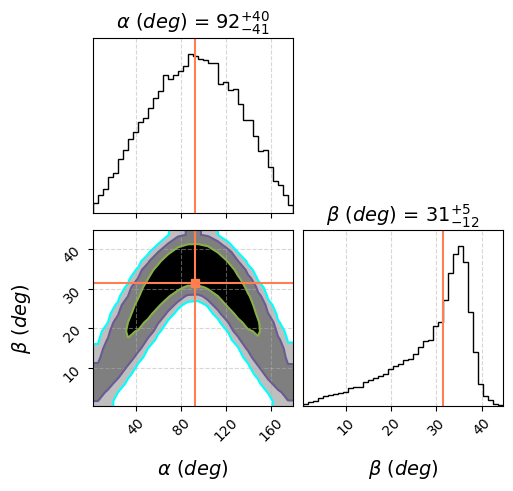}
\caption{Posterior distribution of RVM fit}
\label{fig:rvm}
\end{center}
\end{figure}

$\phi$ represents the pulsar's rotation phase, while $\alpha$ and $\beta$ are the magnetic latitude and sightline impact angle, respectively. $\psi_0$ and $\phi_0$ represent the offsets in PPA and magnetic latitude. Since $\psi_0$ corresponds to the position angle at the steepest gradient point in the PPA swing, we estimated the offsets by maximizing the rate of change of the PPA. Our derived $\psi_0$ and $\phi_0$ values are $34^{\circ}$ and $176^{\circ}$, respectively. We then used Markov Chain Monte Carlo to explore the posterior distributions of $\alpha$ and $\beta$. The median values with 1$\sigma$ errors of the posterior distributions for $\alpha$ and $\beta$ are $92^{\circ +40^{\circ}}_{-41^{\circ}}$ and $31^{\circ +5^{\circ}}_{-12^{\circ}}$, respectively. The banana-shaped joint posterior distribution (Figure \ref{fig:rvm}) indicates the intrinsic degeneracy of the angles in the fitted model.

\subsection{Mass loss} \label{Mass loss rate}
Following the assumptions by \citet{1994ApJ...422..304T}, we consider material flowing radially outward from the companion within the eclipse region, confined to an approximately spherical region centered on the companion with a diameter equal to the eclipse width, 2$R_{E}$, where $R_{E}$ is the eclipse radius. This material initially leaves the companion's surface at escape velocity, which is significantly lower than the pulsar wind velocity ($V_{W}$), and is thus carried away by the pulsar wind with velocity $V_{W}$.
Therefore, the mass-loss rate from the companion can be estimated as,
$\dot{M_{c}}\sim \pi R_{E}^{2} n_{e} m_{p} V_{W} $. 
If the momentum flux of the ablated material is taken to be equal
to the momentum flux of the pulsar wind at the distance of the
companion, then $V_{W}$ = $(U_{E}/n_{e}m_{p})^{1/2}$. Assuming the pulsar wind to be isotropic then we find the energy density of the wind at the companion distance,  $U_{E}$=$(\frac{\dot{E}}{4\pi c a^{2}})$, where $\dot{E}$=$\frac{4\pi^{2}I\dot{P}}{P^{3}}$ is the spin-down energy loss rate of the pulsar and $a=a_{p}+a_{c}$ is the orbital separation between the pulsar and the companion, where, $a_{p}$sin$i$, projected semi-major axis of pulsar orbit $=$ 0.116895 lt-s \citep{2021ApJ...909....6D} and $a_{c}$, semi-major axis of the companion's orbit $=$ $a_{p}$$M_{psr}$/$M_{c}$. 

For PSR J1908$+$2105, $\dot{E}$ is estimated to be approximately 6 $\times$ 10$^{35}$ erg s$^{-1}$, assuming a pulsar moment of inertia of $I = 10^{45}$ g cm$^{2}$ and a separation of about 1.30 $R_{\odot}$, with an inclination angle ($i$) of 90$^{\circ}$. 

At 400 MHz, the eclipse width, $\Delta\phi\sim$ 0.38, based on the orbital phase coverage during the eclipse (see Table \ref{tab:2}). From the DM distribution discussed earlier, we estimate the electron column density near the egress to be $\sim$ 8 $\times$ 10$^{16}$ cm$^{-2}$ (Equation \ref{eq:ne_ex}). This gives an electron volume density of approximately 4 $\times$ 10$^{5}$ cm$^{-3}$, leading to an estimated mass-loss rate from the companion of $\dot{M_{c}} \sim$ 6 $\times$ 10$^{-12}$ $M_{\odot}$ $yr^{-1}$.

At 650 MHz, the eclipse width is reduced to about $\Delta\phi\sim$ 0.31, with an electron column density near the egress of $\sim$ 1.17 $\times$ 10$^{17}$ cm$^{-2}$. The corresponding electron volume density is approximately 7 $\times$ 10$^{5}$ cm$^{-3}$, and the estimated mass-loss rate is $\dot{M_{c}} \sim$ 5.4 $\times$ 10$^{-12}$ $M_{\odot}$ $yr^{-1}$.

At 2368 MHz, the eclipse width is further reduced to around $\Delta\phi\sim$ 0.27, and the electron column density near the egress is $\sim$ 1.8 $\times$ 10$^{18}$ cm$^{-2}$. This results in an electron volume density of approximately 1.1 $\times$ 10$^{7}$ cm$^{-3}$, with an estimated mass-loss rate of $\dot{M_{c}} \sim$ 2 $\times$ 10$^{-11}$ $M_{\odot}$ $yr^{-1}$.

With the mass loss rate at 2368 MHz, the companion can completely evaporate in a 3 Gyr time scale. However, as mentioned by \cite{2018MNRAS.476.1968P}, long-term orbital dynamics will likely affect the evolution of mass loss. If the magnetic braking of the companion and gravitational radiation are negligible, the mass loss will cause the companion to move far away from the pulsar. Combined with the pulsar's spin-down, this decreased irradiation of the companion star over time is expected to reduce the likelihood of complete evaporation.

\subsection{Orbital phase resolved flux density and plasma lensing} \label{flux density}

\begin{figure*}[ht!]
\begin{center}
    \begin{minipage}{0.60\textwidth}
        \centering
        \includegraphics[width=\textwidth]{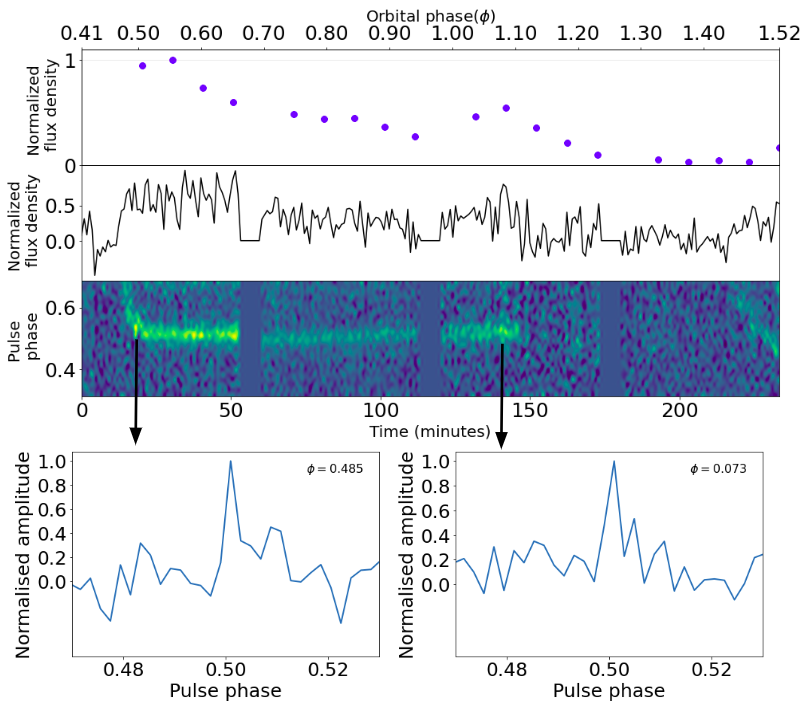}
    \end{minipage}
    \hspace{0.1 cm}
    \begin{minipage}{0.35\textwidth}
        \centering
        \includegraphics[width=\textwidth]{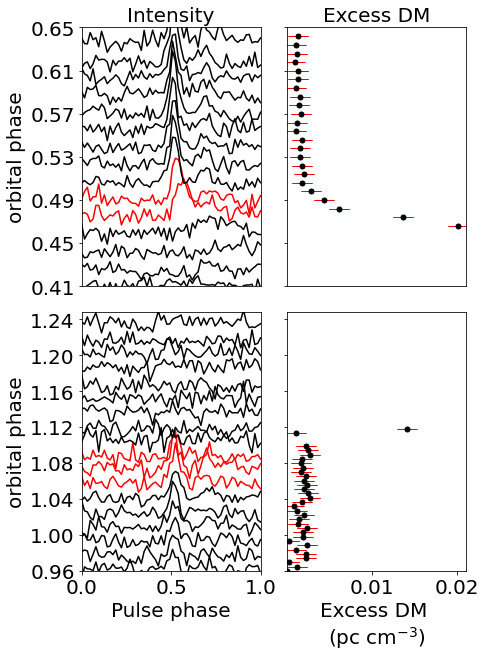}
    \end{minipage}
    \caption{Left panel: Enhanced flux density and increase in brightness in pulsed radiation near eclipse boundaries observed with uGMRT at 400 MHz on 16 Dec 2023 in an observation covering nearly one full orbit ($\sim$ 3.5 hours) of the pulsar, monitoring the MSP starting from one egress and ending at the next egress.}
    Right panel: Total intensity, I (black) for average pulse profiles of PSR J1908$+$2105 during the ingress (bottom panels) and egress (upper panels) of the eclipse from uGMRT observations at 400 MHz. The red lines are the last and first pulses before and after the eclipse. The pulse profiles near the eclipse become wider. The corresponding DM (black dots) variations of the eclipse are shown in the right panels. The red bars are the errors on the DM.
    \label{fig:lensing-flux}
\end{center}
\end{figure*}

The total intensity vs orbital phase is shown in Figure \ref{fig:lensing-flux} for observations at 400 MHz and the first panel of Figure \ref{fig:flux-pol} shows the flux density variation at 2368 MHz. Similar to PSR J2051$-$0827 \citep{https://doi.org/10.1093/mnras/stab1811}, we have seen an enhanced flux density and increase in brightness in pulsed radiation near the first eclipse boundary ($\phi\sim$ 0.45$-$0.60) on epoch 16 December 2023 as shown in Figure \ref{fig:lensing-flux}. The flux density during the eclipse phase of the first eclipse is about four times greater than the flux density observed during the non-eclipse phase. Additionally, single bright pulses were detected just before and after the radio eclipse, particularly with a high occurrence during the egress of the first eclipse. However, no such bright pulses were detected in the non-eclipse region. To determine if this magnification is due to scintillation from the interstellar medium (ISM), we calculated the decorrelation bandwidth ($\approx$ $\frac{1}{2\pi\tau}$) to be 12 kHz. This value is smaller than the frequency resolution (390 kHz) of our data set, indicating that scintillation from the ISM is unlikely to be responsible for the observed magnification. Therefore, we interpret this as lensing due to the irregular distribution of the circumstellar plasma plasma around the companion. In regions characterized by strongly varying electron density, radio emission can undergo significant magnification due to plasma lensing. Plasma lensing was observed at only one of the epochs at 400 MHz. There were no simultaneous observations at other frequency bands at that epoch. We did not detect any lensing effects in our other observations at 400 MHz as well as at 650 and 2368 MHz. We note that multiple plasma lenses may or may not form within a compact region capable of focusing at vastly different frequencies due to the frequency-dependent scaling of dispersive and geometric delays \citep{https://doi.org/10.1038/s41586-018-0133-z}. Moreover, the turbulence and density gradients in the eclipse medium responsible for lensing are random and can vary widely between the observing epochs. Instances of extreme plasma lensing have been observed surrounding eclipses in two other eclipsing spider pulsars: the black widow MSP B1957$+$20 \citep{https://doi.org/10.1038/s41586-018-0133-z}, and the redback MSP B1744$-$24A \citep{2019ApJ...877..125B}. These lensing effects result in highly magnified pulses, amplifying their intensity by up to ten times compared to those from the non-eclipse region, with this enhancement lasting for tens of milliseconds. In PSR B1957$+$20, \citet{https://doi.org/10.1038/s41586-018-0133-z} reported a magnification factor of nearly 40 during lensing events. Similarly, \citet{2019ApJ...877..125B} observed an amplification factor of 10 compared to the pulses from the noneclipse region in PSR B1744$-$24A. Unlike \citet{https://doi.org/10.1038/s41586-018-0133-z}, we do not observe any bright structures in the dynamic spectra. This may be due to the presence of multiple lenses in the eclipse outflow creating an interference pattern among the lensing images and dissolving any distinct features. The lensing phenomenon resolves the pulse emission, thereby constraining emission sizes and component separations. It can also be used to infer the velocity of the eclipsing outflow. 
Following \citet{https://doi.org/10.1038/s41586-018-0133-z}, \citet{2019ApJ...877..125B} and \citet{https://doi.org/10.3847/1538-4357/aa74da}, we tried to constrain the characteristic size of a 1D lens. For $\nu$ =400 MHz, the Fresnel scale at the lens plane is found to be:
\begin{equation}
    r_{F} \approx \sqrt{\frac{cd_{sl}d_{lo}}{\nu d_{so}}} \approx  26 \times \sqrt{d} \; \;\text{km}
\end{equation}
where, we assume, $d_{sl}$ is the distance to the lens from source  $\approx$ d $\times$ a ($a \approx$ orbital separation between the pulsar and the companion $\approx$ 1.30 $R_{\odot}$); $d_{lo}$, distance to the lens from the observer $\approx$ $d_{so}$, distance to the source from the observer $\approx$ 2.6 kpc.
We define the size of the lens, $R_{lens}$ $\approx$ $R$ $\times$ $a$. From \citet{https://doi.org/10.3847/1538-4357/aa74da},  we find the amplification due to lensing,
\begin{equation}
    G \sim \frac{R_{lens}}{r_{F}} \approx 3.4 \times 10^{4}\frac{R}{\sqrt{d}}
\end{equation}
As the maximum gain, $G \approx 4$,  
\begin{equation}
   \frac{R}{\sqrt{d}} \approx 1.15 \times 10^{-4}
\end{equation}

The time of caustic crossing (Equation (22) of \citet{https://doi.org/10.3847/1538-4357/aa74da}) is given by
\begin{equation}
    t_{c} \sim \frac{R_{lens}{\delta G/ G}}{v_{trans}G^{2}}\left (\frac{d_{lo}}{d_{so}} \right ) \approx 132 \times R \;\; \text{s} 
\end{equation}
where, $v_{trans}$ is the relative velocity between the pulsar and companion outflow and can be approximated to the orbital velocity of the companion $\approx$ 428 km s$^{-1}$ \citep{1991ApJ...381L..21T,1993A&A...267L...1T}. We assume the fractional gain $\delta G/G \approx 4/4 \approx 1$. Therefore, for $t_{c}$ $\sim$ $1-50$ ms, we obtain $R$ $\approx$ $7.5 \times 10^{-6}$ $-$ $3.7 \times 10^{-4}$ km. Hence, corresponding $R_{lens}$ $\approx$ 6.8$-$343 km and $d$ $\approx$ $0.004-10.4$. From the obtained value of $d$, we can assume the lens material is located very near to the companion that is at an orbital separation $a$. Therefore, the resolution of the lens for given magnification G $\sim$ $1.9R_{1}/G^{1/2}$ for the linear lens and $1.9R_{1}/G^{1/4}$ for a circular lens \citep{https://doi.org/10.1038/s41586-018-0133-z}; where $R_{1}$ = $\sqrt{\lambda a/\pi}$ $\approx$ $15$ km for our observing length $\lambda = 75 $ cm.
The resulting resolution is 14.3 km for the linear lens and 20 km for the circular lens, both of which are significantly smaller than the light-cylinder radius of PSR J1908+2105 ($R_{LC}$ = $cP/2\pi$ $\approx$ 122 km). This makes lensing highly effective for probing the emission geometry.

\section{Eclipse mechanisms} \label{sec:eclipse mec}

In the following section, we have applied the eclipse mechanism proposed by \citet{1994ApJ...422..304T} to PSR J1908$+$2105 to explore the potential causes of the observed eclipsing. Figure \ref{fig:lensing-flux} represents the total intensity and excess DM near the eclipse ingress and egress at 400 MHz. The eclipse is centered around orbital phase 0.28 at this frequency with a duration of $\sim$ 80 minutes, i.e. nearly 38\% of the orbit. The radius of the
companion’s Roche lobe, $R_{L}$ is estimated by assuming a binary inclination of 90$^{\circ}$ and using the following equation from \cite{1983ApJ...268..368E},
\begin{equation}
R_{L}=\frac{0.49aq^{2/3}}{0.6q^{2/3}+ln(1+q^{2/3})} \sim 0.21 R_{\odot},
\end{equation}
where the mass ratio of the companion to the pulsar, denoted as q = m$_{c}$/m$_{p}$, and the separation between them, a ($\sim$ 1.3 R$_{\odot}$).
The obscured portion of the companion's orbit extends to 1.48 $R_{\odot}$, significantly larger than that of the Roche lobe radius. This indicates that the material causing the eclipse is located well outside the companion's Roche lobe, implying it is not gravitationally bound to the companion.

Since the eclipse weakens at higher frequencies, we can probe deeper into the eclipse medium at these frequencies. We found the maximum electron column density at 2368 MHz to be N$_{e}$ $\sim$ 1.78 $\times$ 10$^{18}$  cm$^{−2}$. However, due to the complete eclipse, only a lower limit on the electron column density could be determined near the egress. From this, we derive an electron volume density of n$_{e}$ $\sim$ 1.1 $\times$ 10$^{7}$ cm$^{−3}$, where n$_{e}$= N$_{e}/L$, and $L$ is the absorption length of eclipse medium at 2368 MHz $\sim$ 2.3 R$_{\odot}$.

We have ruled out plasma frequency cutoff as the primary eclipse mechanism, as the plasma frequency is determined to be $f_{p}$= 8.5($\frac{n_{e}}{cm^{-3}}$)$^{1/2}$ kHz $\sim$ 28.35 MHz, significantly lower than the observed cutoff frequency ($>$ 4000 MHz). Refraction is also excluded as a major mechanism since refraction would require pulse delays in the range of 10$-$100 ms, whereas we observed a delay of 445 $\mu$s at 2368 MHz near the eclipse boundary.

A scattering of radio waves could cause an eclipse if the pulse broadens beyond the pulsar period. However, the scattering timescales at 1200 and 400 MHz (Section \ref{Profile evolution}) are significantly shorter than the pulsar spin period ($<$ 2.56 ms). We found that the scattering time scales are similar in the eclipse phase and the non-eclipse phase. Observed scattering time scale near eclipse ingress is found to be 0.09 and 0.06 ms at 400 MHz and 2 GHz respectively; whereas in the non-eclipse phase, we calculate a scattering time scale of 0.08 ms and 0.05 ms, making scattering an unlikely explanation for the observed eclipsing. 

We also considered whether free-free absorption could account for frequency-dependent eclipsing. For free-free absorption to be the cause, the optical depth $\tau_{ff}$ (given by Equation (11) of \citet{1994ApJ...422..304T}) must exceed 1. Using $\tau_{ff} > 1$, we derived the condition T $\leq$  42 $\times$ $f_{cl}^{2/3}$ K, where $T$ is the temperature and $f_{cl}$ is the clumping factor. For significant free-free absorption across both high and low frequencies, either a very low temperature or a very high clumping factor would be required. However, neither condition is physically feasible: the temperature of stellar wind (10$^{8}$-10$^{9}$ K) is well above the necessary threshold, and an extremely high clumping factor is not plausible in the eclipse environment \citep{1994ApJ...422..304T}.

Induced Compton scattering is also ruled out as a major eclipse mechanism, as the calculated optical depth (using equation (26) from \citet{1994ApJ...422..304T}) is less than 1. The optical depth for induced Compton scattering, estimated with a mean flux density of 0.22 $\pm$ 0.01 mJy, a pulsar spectral index of $\alpha$=$−$1.9, a companion distance of approximately 1.3 solar radii, and a pulsar distance of 3.05 kpc (derived from the DM of the pulsar using NE2000 and YMW16 models, averaging the estimates of 2.6 kpc and 3.2 kpc), is too low to account for the observed eclipse.

We also explored whether cyclotron absorption could explain the observed eclipse.
We calculated the characteristic magnetic field (B$_{E}$) in the eclipse medium to be 16 G,  The calculated cyclotron frequency, $\nu_{B}=\frac{eB}{2\pi m_{e}c}$ is 44 MHz; where $m_{e}$ is the mass of the electron, $e$ is the charge on the electron and $c$ is the speed of light. Therefore, the corresponding cyclotron harmonic, $m$, at the observing frequency $\nu$ can be calculated using $m=\frac{\nu}{\nu_{B}}$. According to Equation (43) from \citet{1994ApJ...422..304T}, the temperature within the eclipse medium needs to be approximately 1.5 $\times$ 10$^{8}$ K for cyclotron absorption to occur at 400 MHz. However, the validity of the cyclotron approximation holds for temperatures T $\leq$ 3.9 $\times$ 10$^{6}$ K at 400 MHz. At a higher frequency of 2368 GHz, the required temperature for cyclotron absorption reduces to 6.1 $\times$ 10$^{7}$ K but cyclotron approximation is valid for only T $\leq$ 1.9 $\times$ 10$^{4}$ K at 2368 MHz. Thus, cyclotron absorption is not the predominant mechanism explaining the observed eclipses across all frequencies.

\begin{figure*}[ht!]
\begin{center}
    \begin{minipage}{0.45\textwidth}
        \centering
        \includegraphics[width=\textwidth]{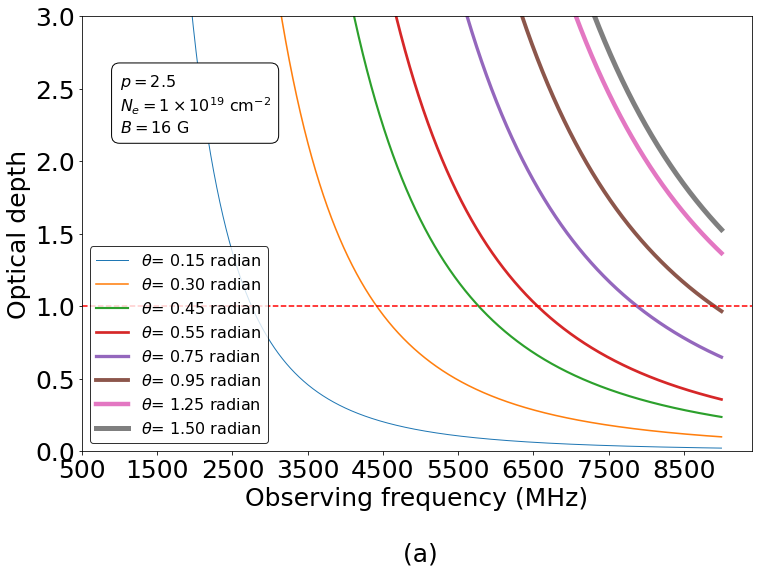}  
    \end{minipage}
    \hspace{0.5 cm}
    \begin{minipage}{0.45\textwidth}
        \centering
        \includegraphics[width=\textwidth]{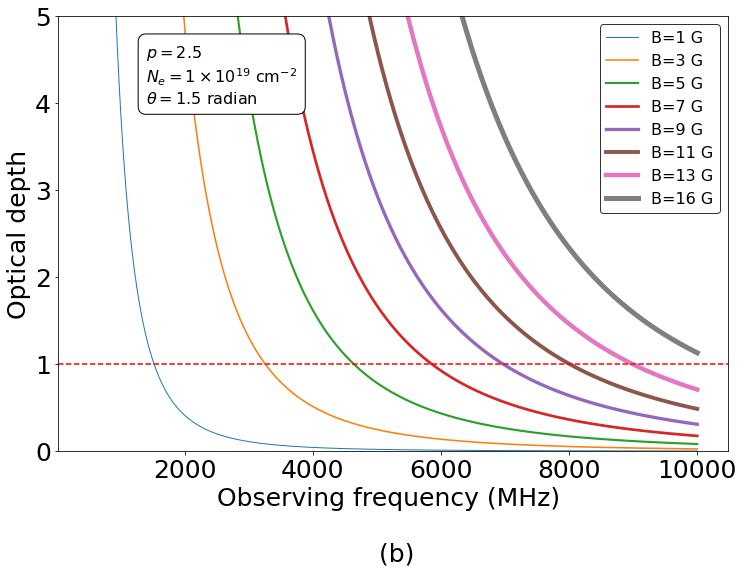}       
    \end{minipage}
\end{center}
\begin{center}
    \begin{minipage}{0.45\textwidth}
        \centering
        \includegraphics[width=\textwidth]{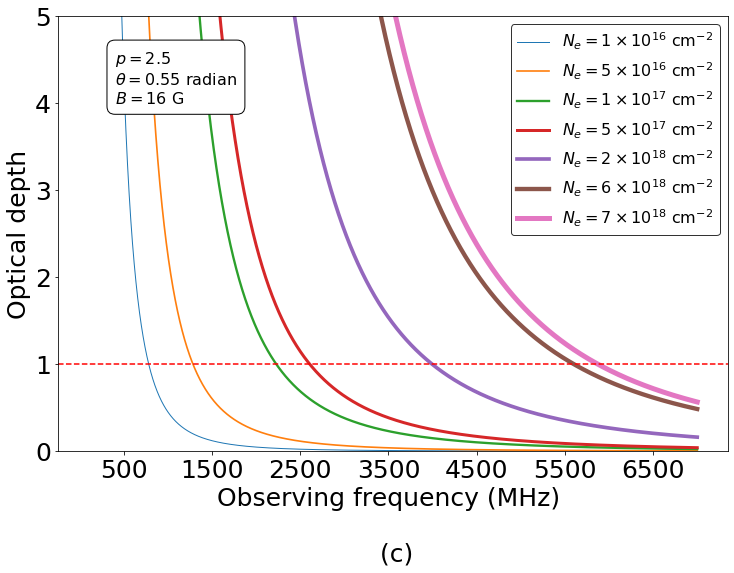}       
    \end{minipage}
    \hspace{0.5 cm}
    \begin{minipage}{0.45\textwidth}
        \centering
        \includegraphics[width=\textwidth]{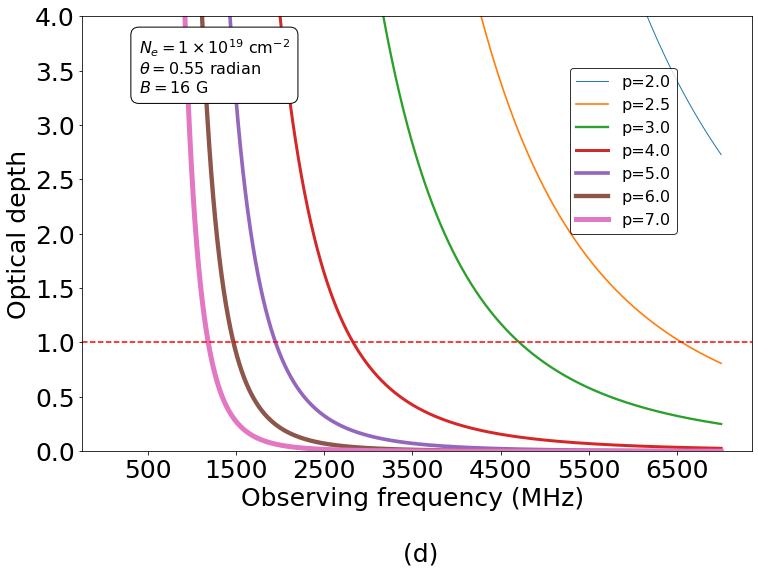}       
    \end{minipage}
\end{center}
\caption{Simulation of optical depth variation with frequency through changes in viewing angle $(\theta)$ (a), magnetic field $(B)$ (b), electron density $(N_{e})$ (c), and electron power law index $(p)$ (d) in the eclipse medium for PSR J1908$+$2105. The axis ranges were selected to enhance the plot's clarity and provide a better explanation of optical depth variation as we vary different parameters.}
\label{fig:sync}
\end{figure*}

% \usepackage{booktabs}
\renewcommand{\tabcolsep}{5pt} % Adjust global column spacing
\begin{table*}[!htb]
\begin{center}
{\footnotesize
\caption{Maximum Electron column density and Eclipse cutoff frequency for spider MSPs}
\label{tab:3}
\begin{tabular}{lccc@{\hskip 10pt}c} % Add extra spacing between the last two columns
\hline
\toprule
MSP & $N_{e}$ maximum & cutoff frequency ($\nu_{c}$) & $\tau^{\ddag}_{ind}$ & References \\ 
& ($cm^{-2}$) & (MHz) &  \\
\midrule
\hline
J0024$−$7204J$^{*}$   &  5 $\times$ 10$^{16}$ & $<$ 734 & 1.29 $\times$ 10$^{-3}$ &(6)\\
J1023$+$0038$^{*}$    & 4.6 $\times$ 10$^{17}$ & $\sim$ 3000 & 1.12 $\times$ 10$^{-4}$ & (1), (2) \\
J1227$–$4853$^{\dag}$ & 2.4 $\times$ 10$^{17}$ & 750 $<\nu_{c}<$ 1400 & 7.6 $\times$ 10$^{−5}$ & (7) \\
J1431$−$4715$^{*}$    & 9.0 $\times$ 10$^{17}$ & $\sim$ 1251 & 1.98 $\times$ 10$^{-5}$ & (6) \\
J1544$+$4937$^{*}$    &  3.0 $\times$ 10$^{16}$ & 338 $<\nu_{c}<$ 558 & 4 $\times$ 10$^{-2}$  & (5)\\
{\bf J1731$−$1847}$^{\dag}$ & 3.0 $\times$ 10$^{18}$ & $>$ 3000 & 3.96 $\times$ 10$^{-4}$ & (2) \\
J1748$−$2446A$^{*}$   &  2.0 $\times$ 10$^{18}$ & 1600 $<\nu_{c}<$ 2600 & 7.90 $\times$ 10$^{-3}$ & (2),(3)\\
J1810$+$1744$^{\dag}$   &  3.0 $\times$ 10$^{16}$ & $>$850$^{a}$ & 6.66 $\times$ 10$^{-4}$ & (8),(11)\\
J1816$+$4510$^{\dag}$   &  3.0 $\times$ 10$^{17}$ & $>$1400$^{a}$ & 4.94 $\times$ 10$^{-5}$ & (8),(9)\\
{\bf J1908$+$2105}$^{\dag}$ & 2.0 $\times$ 10$^{18}$ & $>$ 4000 & 1.85 $\times$ 10$^{-3}$ & This work \\
%B1957$+$20$^{\dag}$   &  4.0 $\times$ 10$^{17}$ & $>$1400$^{a}$ & (2), (10)\\
J1959$+$2048$^{\dag}$ &  4.0 $\times$ 10$^{17}$ & $>$1400$^{a}$ & 6.43 $\times$ 10$^{-5}$ & (2),(6),(10)\\
J2051$−$0827$^{*}$    &  4.0 $\times$ 10$^{17}$ & $\sim$ 1000 & 5.91 $\times$ 10$^{-4}$ &(2),(4)\\
\hline
\end{tabular}
}
\end{center}

{\footnotesize {{\bf Notes.}}\\
MSPs listed in bold are those that exhibit eclipses at frequencies above 3 GHz.\\
$^{*}$ Electron density at the center of the eclipse medium.\\
$^{\dag}$ Electron density at the eclipse boundary.\\
$^{a}$ Cutoff frequency is not reported due to band-limited spectra above 1.4 GHz.\\
$^{\ddag}$ Optical depth for Compton scattering at the cutoff frequency was calculated using equation (26) from \citet{1994ApJ...422..304T} using the parameters from the corresponding references and ATNF pulsar catalog \citep{2005AJ....129.1993M}. 

{\bf{refs:}} 1.\citet{2009Sci...324.1411A}, 2.\citet{https://doi.org/10.1111/j.1365-2966.2011.18416.x}, 3.\citet{2018ApJ...867...22Y}, 4.\citet{ https://doi.org/10.1046/j.1365-8711.2001.04074.x}, 5.\citet{https://doi.org/10.3847/1538-4357/ad0b83}, 6.\citet{https://doi.org/10.48550/arXiv.2407.01024}, 7.\citet{https://doi.org/10.3847/1538-4357/aba902}, 8.\citet{ https://doi.org/10.1093/mnras/staa596}, 9.\citet{http://dx.doi.org/10.1088/0004-637X/791/1/67}, 10.\citet{1991ApJ...380..557R}, 11.\citet{2018MNRAS.476.1968P}}
\end{table*}

However, according to \citet{1994ApJ...422..304T}, absorption by thermal electrons dominates at lower cyclotron harmonics. As we move toward the higher harmonic, synchrotron absorption by the transrelativistic
nonthermal free electrons become significant and therefore it can be thought of as a major eclipse mechanism at higher frequencies.
\citet{https://doi.org/10.3847/1538-4357/ad0b83}, \citet{https://doi.org/10.3847/1538-4357/ac19b9} found that synchrotron absorption can also explain the frequency-dependent nature of eclipse. Therefore, we investigated the possible mechanism for the eclipse observed at unusually high frequencies for PSR J1908$+$2105, focusing on synchrotron absorption as the major eclipse mechanism. Our goal was to determine what is different in the eclipse medium of this system compared to other known spiders, leading to such high-frequency eclipses. 
The energy density distribution of the nonthermal electrons for synchrotron absorption is, $(n(E) = n_{0}E^{−p})$, where $p$ is the power-law index and $n_{o}$  is the nonthermal electron density which is assumed to be 1\% of the total electron density ($n_{e}$). The optical depth for synchrotron absorption is given by:
\begin{equation}
\tau_{syn}=\left(\frac{3^{\frac{(p+2)}{2}}\Gamma(\frac{3p+2}{12}) \Gamma (\frac{3p+22}{12})}{4} \right) \left(\frac{sin\theta }{m}\right)^{\frac{p+2}{2}} \frac{n_{0}e^{2}}{m_{e}c\nu}L
\end{equation}
where $\theta$ is the angle between the magnetic field lines at the eclipse medium and our line of sight, and $L$ is the absorption length. 
Figure (\ref{fig:sync}) illustrates the optical depth in the eclipse medium depending on the parameters $p$, $\theta$, $N_{e}$, and $B$. We can see, that the optical depth and, therefore the eclipse cutoff frequency can vary if we change any of these four parameters\footnote{\url{https://github.com/Rimi98/eclipse_mechanism}}, and therefore, constraining these parameters accurately for a particular system by removing the degeneracy is almost impossible. However, the observed eclipse at a frequency $>4$ GHz for PSR J1908$+$2105, helps us to get a better constrain and put a limit on these parameters, $N_{e}$, $B$,$p$, and $\theta$. In Table (\ref{tab:3}), we have compared the maximum electron density and corresponding eclipse cutoff frequency found for a few of the spider systems, for which a detailed eclipse study has been done. MSPs exhibiting eclipses at higher frequencies tend to have higher electron densities compared to those with lower eclipse cutoff frequencies. Notably, MSPs that exhibit eclipses at frequencies greater than 3 GHz have an electron density ($N_{e}$) of approximately $10^{18}$ cm$^{-2}$ at the eclipse boundary, which is an order of magnitude higher than the electron density found at the center of the eclipse medium for other eclipsing spiders. 
Therefore, higher electron density in the eclipse medium can be one reason for the observed high-frequency eclipse for PSR J1908$+$2105. Thus we infer higher electron column density can play a major role in deciding the eclipse cutoff frequency of frequency-dependent eclipsing.
Optical depth for Compton scattering at the cutoff frequency (also presented in Table \ref{tab:3}) ranges from $10^{-5}$ to $10^{-2}$, indicating that this mechanism may not be a major contributor to eclipsing at higher frequencies.

For absorption by nonthermal electrons to occur, the electron power-law index can vary from 2 to 7 for the range of cyclotron harmonic numbers, $m=\frac{\nu}{\nu_{B}}\approx$ 10$-$100 and viewing angle ($\theta$) from 20$^{\circ}$ to 80$^{\circ}$ \citep{1982ApJ...259..350D}. For absorption by nonthermal, mildly relativistic electrons this value of electron power law index converges between 2 to 3 as shown with the numerical calculation by \citet{1970SoPh...11..434T,1969ApJ...158..753R}
Therefore, to constrain the viewing angle $\theta$ for our observed high-frequency eclipse, we have kept the $p=2.5$, and taken the magnetic field in the eclipse medium $B=16$ G as we obtained in Section \ref{Magnetic Field}, and predicted electron density at the eclipse medium $N_{e}=1\times10^{19}$ and varied $\theta$ from 0.15 radian to 1.50 radian \citep{https://doi.org/10.3847/1538-4357/ac19b9}. It can be seen in panel (a) of Figure (\ref{fig:sync}) that, $\theta$ should be $>0.30$ radian for the eclipse to occur at $>$ 4 GHz.

Next, we set $p = 2.5$ and used the predicted maximum electron column density at the eclipse center,  $N_{e} = 1 \times 10^{19}$ cm$^{-2}$. Assuming the magnetic field is perpendicular to the line of sight (i.e., $\theta$ = 1.5 radian), we varied the magnetic field ($B$) from 1 Gauss to 16 Gauss. This analysis suggests that the lower limit of the magnetic field in the eclipse medium must be $\geq$ 5 Gauss to account for the observed eclipse (see panel (b), Figure \ref{fig:sync}).

Although it is already evident, we can see in panel (c) of Figure \ref{fig:sync} that, keeping $p=2.5$, $B=16$ Gauss and $\theta=0.55$ radian fixed, the eclipse cutoff frequency shifts towards higher frequency as we increase the electron density.

Lastly, keeping $N_{e}=1\times10^{19}$ cm$^{-2}$, B=16 Gauss and $\theta$=0.55 radian fixed, we can see for the eclipse to occur at frequency $>$ 4 GHz, the electron power law index ($p$) for synchrotron absorption should be $3<p<2$ (panel (d), Figure \ref{fig:sync}) suggesting major eclipse mechanism to be absorption by non-thermal transrelativistic electrons \citep{1970SoPh...11..434T,1969ApJ...158..753R}.

Using a similar approach and considering N$_{e}$ reported, we determined that the minimum magnetic field required to explain the observed eclipse cutoff at 3000 GHz for PSR J1023$+$0038 is 10 Gauss, with an upper limit of 200 Gauss. For PSR J1431$-$4715, the minimum magnetic field needed for an eclipse cutoff at 1.2 GHz is 2 Gauss, with a maximum limit of 35 Gauss. Similarly, for PSR J2051$-$0827, the required magnetic field for the eclipse cutoff at 1 GHz ranges from a minimum limit of 2 Gauss to a maximum limit of 50 Gauss.

Therefore, we have inferred that synchrotron absorption by free electrons is the primary mechanism responsible for the observed eclipses in PSR J1908$+$2105.

\section{summary} \label{sec:summary}
This study of the spider MSP J1908$+$2105 significantly advances our understanding of the magnetic properties within the eclipse medium of spider MSPs, shedding light on the dynamics of these extreme binary systems. PSR J1908$+$2105 stands out as a unique system that blurs the lines between ``black widows" and ``redbacks." Its short orbital period, low companion mass, and extensive eclipses characterize it as an atypical "spider" variant, sharing traits of both subclasses. Notably, it exhibits unusual eclipses over 30\% of its orbit at frequencies exceeding 4 GHz. The other two MSPs exhibiting eclipses above 3 GHz are PSR 1723$-$2837 and PSR 1731$-$1847. PSR 1723$-$2837, which falls into the redback category, has a companion mass ranging from 0.4 to 0.7 $M_{\odot}$. PSR 1731$-$1847, with a companion mass of 0.043 $M_{\odot}$, is classified as a black widow.

We explain the trailing material observed at the eclipse boundaries by modeling the electron density distribution around companion stars. The linear depolarization and increased RM near the eclipse medium suggest a higher electron density and magnetic field, approaching closer to the companion. While extensive eclipses above 4 GHz have been observed in three spider MSPs, detailed studies for the other two are lacking. 

For PSR J1908$+$2105, our wide-bandwidth observations identify synchrotron absorption as the primary eclipse mechanism, allowing us to establish a minimum required magnetic field for eclipses at such high frequencies. We also found a correlation between eclipse cutoff at higher frequencies and increased electron density. The emission geometry, constrained from the observed position angle (PPA) swing, has been characterized in only a few spider MSPs.

Additionally, we observed lensing effects in uGMRT band$-$3 near eclipse boundaries, causing enhanced flux density and brightness. 
These findings provide crucial insights into the complex interactions within spider MSP systems, offering a more comprehensive understanding of the eclipse mechanisms and magnetic environments that define these intriguing binaries.
\par 

We acknowledge the support of the Department of Atomic Energy, Government of India, under project No.12-R\&D-TFR5.02-0700. The GMRT is run by the National Centre for Radio Astrophysics of the Tata Institute of Fundamental Research, India. We acknowledge the support of GMRT telescope operators for observations. The Parkes radio telescope is part of the Australia Telescope which is funded by the Commonwealth of Australia for operation as a National Facility managed by CSIRO. We sincerely thank the reviewer for valuable feedback and insightful suggestions that improved the clarity and presentation of this work.

\bibliography{J1908.bib}{}
\bibliographystyle{aasjournal}

\end{document}